\documentclass[preprintnumbers,
nofootinbib,
notitlepage,
amsmath,amssymb,
]{revtex4-1}

\usepackage{graphicx}
\usepackage{dcolumn}
\usepackage{bm}

\makeatletter
\def\slash#1{{\mathpalette\c@ncel{#1}}} 
\makeatother

\newcommand\beq{\begin{eqnarray}}
\newcommand\eeq{\end{eqnarray}}

\newcommand{\Slash}[1]{{\ooalign{\hfil/\hfil\crcr$#1$}}} 

\newcommand\la{\langle}
\newcommand\ra{\rangle}

\newcommand{\tr}{{\rm Tr}  }
\newcommand{\nn}{\nonumber \\}

\begin{document}

\preprint{YITP-16-79}

\title{Single Spin Asymmetry in Forward $pA$ Collisions}


\author{Yoshitaka Hatta$^{\rm a}$}
\author{Bo-Wen Xiao$^{\rm b}$}
\author{Shinsuke Yoshida$^{\rm b}$}%
\author{Feng Yuan$^{\rm c}$}

\affiliation{\vspace{3mm}
  $^{\rm a}$Yukawa Institute for Theoretical Physics, Kyoto University, Kyoto 606-8502, Japan \\
  $^{\rm b}$ Key Laboratory of Quark and Lepton Physics (MOE) and Institute of Particle Physics, Central China Normal University, Wuhan 430079, China \\
  $^{\rm c}$ Nuclear Science Division, Lawrence Berkeley National Laboratory, Berkeley, CA 94720, USA \\
}%

\date{\today}

\begin{abstract}
\vspace{4mm}
We compute the transverse single spin asymmetry in light hadron production $p^\uparrow p\to hX$ and $p^\uparrow A\to hX$  including the gluon saturation effect in the unpolarized nucleon/nucleus. In the forward (large-$x_F$) region, the dominant contribution comes from the so-called derivative term associated with the soft gluonic pole. This leads to the cancellation of nuclear effects in $A_N$ which can be tested at RHIC.  We also show that the soft fermionic pole disappears in the saturation environment.

\end{abstract}

\pacs{Valid PACS appear here}
\maketitle




\section{Introduction}

Recently, growing attention has been given to the interplay between spin physics and small-$x$ physics. 
While the two subjects are usually discussed by different communities, there are interesting mutual 
problems of direct phenomenological importance. For example, the small-$x$/Regge behavior of the polarized 
parton distribution functions $\Delta q(x)$ and $\Delta g(x)$ is  relevant to the nucleon spin decomposition problem 
\cite{Bartels:1995iu,Hatta:2009ra,deFlorian:2014yva,Kovchegov:2015pbl}. 
Also, various single spin asymmetries (SSAs) in $pp$ and $pA$ collisions have been computed by 
including the gluon saturation effects 
\cite{Boer:2006rj,Kang:2011ni,Kovchegov:2012ga,Kang:2012vm,Zhou:2013gsa,Altinoluk:2014oxa,Schafer:2014zea,Zhou:2015ima,Boer:2015pni}. 
On the experimental side, RHIC has recently reported its first measurement of SSA 
on a nuclear target \cite{Heppelmann:2016siw} that might call for a saturation-based explanation. 
More connections of this sort will certainly be explored at the future Electron-Ion Collider (EIC)~\cite{Accardi:2012qut,Boer:2011fh}. 

In this paper, we revisit the transverse SSA  in light hadron production $p^\uparrow p\to hX$ 
or $p^\uparrow A \to h X$. This process has been extensively discussed in the literature in the collinear  twist-three 
approach at high $P_{hT}$ 
\cite{Qiu:1991pp,Qiu:1998ia,Kanazawa:2000hz,Kouvaris:2006zy,Eguchi:2006qz,Eguchi:2006mc,Koike:2007rq,Koike:2009ge,Yuan:2009dw,Kang:2010zzb,Beppu:2010qn,Metz:2012ct,Kanazawa:2013uia,Kanazawa:2014dca} and also, phenomenologically, in the $k_T$-factorization approach at moderate $P_{hT}$ \cite{Sivers:1989cc,Anselmino:1994tv,Anselmino:2005sh}. 
Throughout this paper, we shall focus on the forward rapidity (large-$x_F$) region of the projectile (polarized proton) where SSA is known to be largest. In this region, it is necessary to properly treat the small-$x$ gluons from the target (unpolarized proton/nucleus). In particular, at very high energy and/or for a large nucleus, the saturation effect \cite{Iancu:2003xm} must be taken into account.    
 The first exploratory study in this direction was done    
in \cite{Boer:2006rj} where SSA was given by the convolution of the Sivers function \cite{Sivers:1989cc} 
for the projectile and the unintegrated gluon distribution function for the target including saturation effects. 
Another contribution to SSA in $pA$ collisions from the Collins fragmentation function \cite{Collins:1992kk} 
was calculated in \cite{Kang:2011ni}.

In this work, we employ the `hybrid approach' \cite{Schafer:2014zea} where 
the collinear, twist-three Efremov-Teryaev-Qiu-Sterman (ETQS)  functions \cite{Efremov:1984ip,Qiu:1991pp} 
is used for the projectile and the unintegrated gluon distribution for the target.  The use of the collinear functions instead 
of the ($k_T$-dependent) Sivers function as in \cite{Boer:2006rj} is preferable for a number of reasons. First, the 
$k_T$-dependent factorization is not valid for this process, whereas the hybrid approach has been tested
up to one-loop order for spin-averaged cross sections~\cite{Altinoluk:2011qy,Chirilli:2011km,Chirilli:2012jd,Altinoluk:2014eka,Watanabe:2015tja,Ducloue:2016shw}. Our derivations in this paper
will provide important support to generalize the factorization arguments 
to  spin dependent observables. 
Second, the Sivers function is process-dependent~\cite{Collins:2002kn}, and one
cannot identify the Sivers function used in the phenomenological $k_T$-factorization formula 
with the ones used in the DIS and Drell-Yan processes. 
The collinear twist-three analysis for  the
polarized proton is the appropriate approach to consistently take into account 
the initial and final state interaction effects, which are the key 
components to generate the necessary phase for a non-zero SSA.
Finally, the $k_T$-factorization approach misses important 
contributions to SSA, in particular, the so-called derivative term 
which becomes dominant in the forward region. This term naturally 
arises in our framework and qualitatively changes the behavior of SSA in the forward region.


According to the `hybrid approach', the spin-averaged, inclusive 
hadron production in the forward $pA$ collisions can be written 
as 
\begin{equation}
\frac{d^3\sigma (pA\to hX)}{dy_hd^2P_{hT}}=\int_{x_F}\frac{dz}{z^2}D_{h/q}(z)x_pq(x_p)F(x_g,P_{hT}/z) \ , \label{factori}
\end{equation}
where $y_h$ and $P_{hT}$ are the rapidity and the transverse momentum of the final state hadron, respectively. 
$q(x_p)$ is the collinear quark distribution function and $D(z)$ is the fragmentation function. 
 $F(x_g,k_T)$ is the so-called dipole gluon distribution whose definition will be given in Section \ref{saturation}. In the forward region where $x_g$ is small, $F(x_g)$ includes the saturation effects in the unpolarized target. 
 As mentioned above, the factorization formula (\ref{factori}) has been 
computed up to next-to-leading order in perturbative QCD.

The analog of (\ref{factori}) for the spin-dependent part of the cross section is, schematically, 
\begin{equation}
\frac{d^3\Delta\sigma (p^\uparrow A\to hX)}{dy_hd^2P_{hT}}=\epsilon^{\alpha\beta}P_{h\alpha}S_{T\beta}\int_{x_F}\frac{dz}{z^2}D_{h/q}(z) G_F(x_p,x_p)\otimes F(x_g,P_{hT}/z) \ ,
\end{equation}
where $S_T$ is the traverse polarization vector of the projectile. $G_F(x,x)$ represents the generic twist-three quark-gluon-quark (ETQS) correlation functions which will be defined in Sec.~II.
We shall show that the spin-dependent  cross section can be indeed written in this factorized form 
and clarify the meaning of the symbol $\otimes$.

The rest of this paper is organized as follows. 
In Section II, we quickly review the technique to compute SSA in the collinear factorization framework. In Section III, we calculate the relevant hard matrix elements  in the leading twist approximation and check the consistency with the fully collinear results previously obtained in the  high-$P_{hT}$ region.  We then include the saturation effect in Section IV and discuss the fate of the soft gluonic pole and the soft fermionic pole. At the end we discuss the phenomenological implications of our results. 



\section{Collinear factorization approach}
\label{co}

In the collinear factorization approach, SSA is a twist-three observable which arises from 
multi-parton correlations
in the transversely polarized proton and in the fragmentation process. In this work we shall only consider the former contribution for which the formalism 
to derive the spin-dependent cross section is by now well established 
 \cite{Qiu:1991pp,Qiu:1998ia,Kanazawa:2000hz,Kouvaris:2006zy,Eguchi:2006qz,Eguchi:2006mc,Koike:2007rq,Koike:2009ge,Yuan:2009dw,Kang:2010zzb,Beppu:2010qn,Metz:2012ct,Kanazawa:2013uia}. Here we briefly recapitulate the main steps of the derivation.

SSA in collinear factorization is generated by a series of diagrams shown in Fig. 1. The contribution from the first diagram can be written as 
\beq
\int d^4\xi  d^4\eta\int{d^4k_1d^4k_2\over (2\pi)^8}
e^{ik_1\cdot\xi+i\eta\cdot(k_2-k_1)}
\la pS_{T}|\bar{\psi}_j(0)gA_{\alpha}(\eta)\psi_i(\xi)|pS_{T}\ra
H^{\alpha}_{ji}(k_1,k_2,k,{P_h\over z})\,,
\label{collinear}
\eeq
where $p^\mu\approx \delta^\mu_+ p^+$ is the projectile momentum and  $S_T^\mu$ is the transversely polarized spin four-vector normalized as $S_T^2=-\vec{S}_T^2 =-1$. $i,j$ are the Dirac indices. 
\begin{figure}[h]
\begin{center}
  \includegraphics[height=4cm,width=12cm]{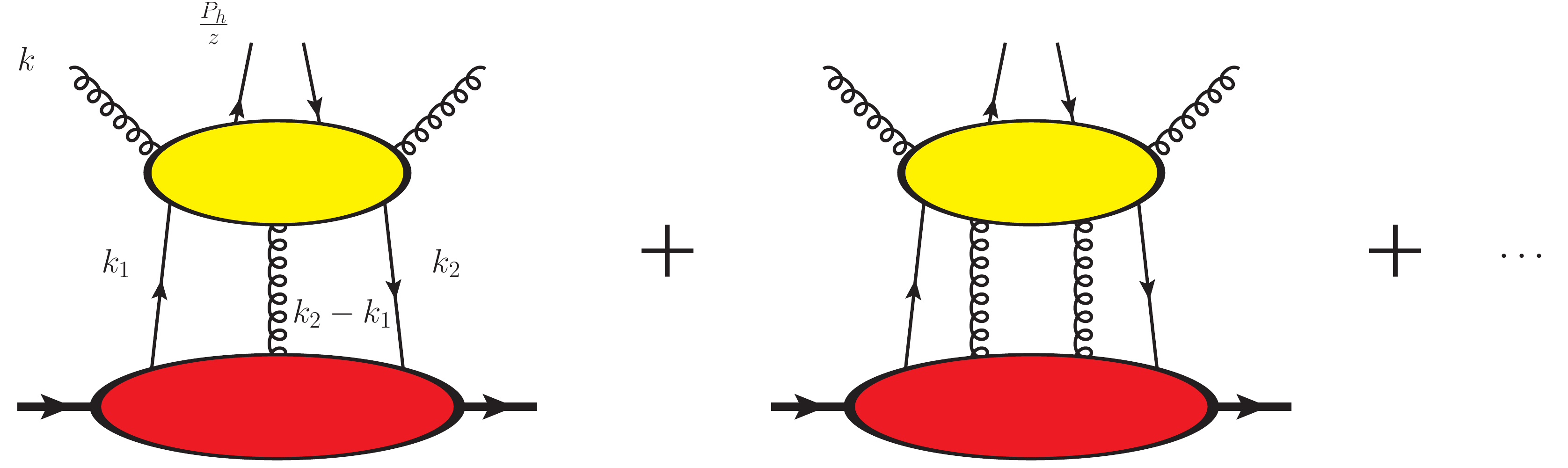}\hspace{1cm}
\end{center}
 \caption{The upper blob represents the hard part $H^{\alpha}_{ji}(k_1,k_2,k,{P_h\over z})$
 and the lower blob
 represents the matrix element of the transversely polarized proton.
 }
\end{figure}
\begin{figure}[h]
\begin{center}
  \includegraphics[height=4cm,width=12cm]{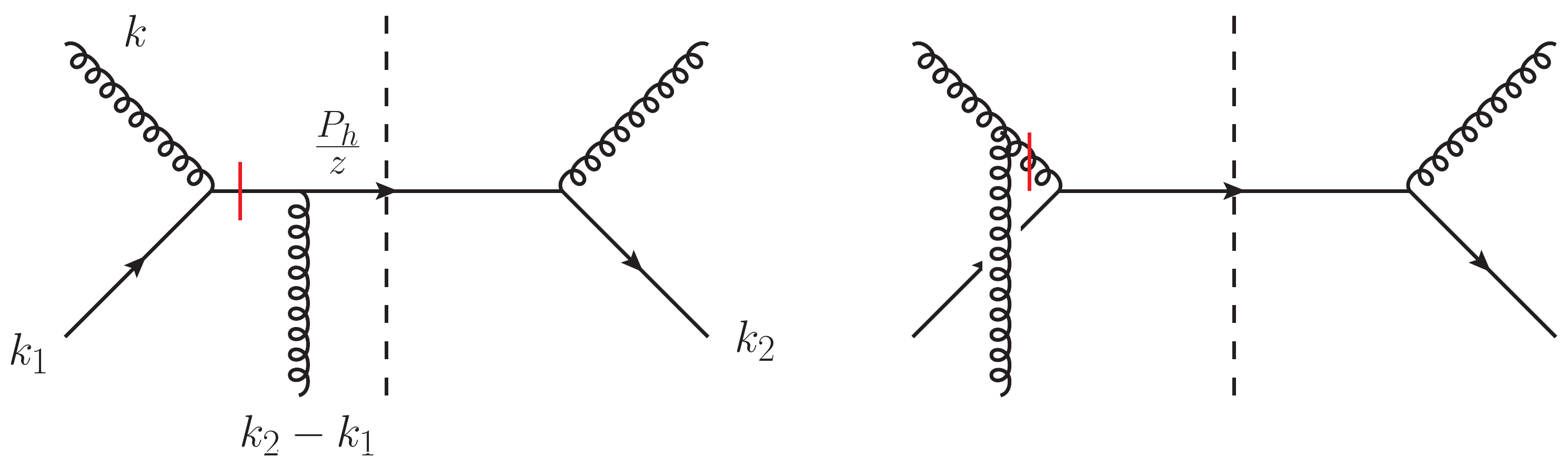}\hspace{1cm}
\end{center}
 \caption{Diagrammatic representation of the hard part $H^{\alpha}_{ji}(k_1,k_2,k,{P_h\over z})$. 
 Barred propagators are `cut' propagators. \label{fig2}
 }
\end{figure}
It is well known that, in order to obtain nonvanishing SSA, one has to pick up the pole of an internal propagator $\frac{i}{p^2+i\epsilon} \to \pi \delta(p^2)$ in  
the hard scattering amplitude $H^{\alpha}_{ji}(k_1,k_2,k,{P_h\over z})$. In the fully collinear calculations \cite{Qiu:1991pp,Qiu:1998ia,Kanazawa:2000hz,Kouvaris:2006zy,Eguchi:2006qz,Eguchi:2006mc,Koike:2007rq}, $H^\alpha_{ji}$ contains a hard $2\to 2$ scattering necessary to produce the transverse momentum $P_{hT}$ of the observed hadron. On the other hand, in our hybrid approach which focuses on the forward region, $P_{hT}$ is provided by the intrinsic transverse momentum of the target. We thus consider the two diagrams in Fig.~2. The barred propagator represents the pole part $\pi \delta (p^2)$, and below we only keep this part in $H^\alpha_{ji}$.  It then satisfies the  Ward identity 
\beq
(k_2-k_1)_{\alpha}H^{\alpha}_{ji}(k_1,k_2,k,{P_h\over z})=0\,.
\label{Ward}
\eeq
From this, it easily follows that ($H^p \equiv H^\mu p_\mu$)
\beq
{\partial\over \partial k_2^{\alpha}}H^{p}_{ji}(k_1,k_2,k,{P_h\over z})\Bigr|_{k_i=x_ip}
=-{\partial\over \partial k_1^{\alpha}}H^{p}_{ji}(k_1,k_2,k,{P_h\over z})\Bigr|_{k_i=x_ip}
={1\over x_1-x_2}H_{\alpha,ji}(x_1p,x_2p,k,{P_h\over z})\,,
\label{SFP}
\eeq
  where $x_1$, $x_2$ are the longitudinal momentum fraction carried by the quarks. In (\ref{SFP}), it is assumed that $x_1\neq x_2$. 

In order to extract the twist-three contribution from (\ref{collinear}), we perform the collinear expansion in the hard part 
\beq
H^{\alpha}_{ji}(k_1,k_2)&=&
H^{\alpha}_{ji}\left(x_1 p,x_2 p\right)
+{\partial\over \partial k_1^{\alpha}}H^{p}_{ji}
(k_1,k_2)\Bigr|_{k_i=x_i p}\omega^{\alpha}_{\ \beta}k_1^{\beta} +{\partial\over \partial k_2^{\alpha}}H^{p}_{ji}
(k_1,k_2)\Bigr|_{k_i=x_i p}\omega^{\alpha}_{\ \beta}k_2^{\beta} \nonumber\\
&=&H^{\alpha}_{ji}(x_1 p,x_2 p)
+{\partial\over \partial k_2^{\alpha}}H^{p}_{ji}
(k_1,k_2)\Bigr|_{k_i=x_i p}\omega^{\alpha}_{\ \beta}(k_2^{\beta}-k_1^{\beta})\,,
\eeq
where $\omega^{\alpha \beta}\equiv g^{\alpha \beta}-\delta^{\alpha}_+ \delta^{\beta}_-$. 
Expanding also the gluon field operator 
\beq
A^{\alpha}={A^+\over p^+}p^{\alpha}+\omega^{\alpha}_{\ \beta}A^{\beta}\,,
\eeq
we obtain
\beq
&&\int d^4\xi  d^4\eta\int{d^4k_1 d^4k_2\over (2\pi)^8}
e^{ik_1\cdot\xi+i\eta\cdot(k_2-k_1)}
\Biggl\{ \la pS_{T}|\bar{\psi}_j(0)g\omega_{\alpha}^{\ \beta}A_{\beta}(\eta)\psi_i(\xi)|pS_{T}\ra
H^{\alpha}_{ji}(x_1p,x_2p) \nn 
&& \qquad \qquad + 
{1\over p^+}\la pS_{T}|\bar{\psi}_j(0)gA^+(\eta)\psi_i(\xi)|pS_{T}\ra {\partial\over \partial k_2^{\alpha}}H^{p}_{ji}(k_1,k_2)\Bigr|_{k_i=x_i p}
\omega^{\alpha}_{\ \beta}(k_2^{\beta}-k_1^{\beta})
 \Biggr\}
\nonumber\\
&&=\frac{i\omega_{\alpha}^{\ \beta}}{p^+} \int dx_1 dx_2
\int \frac{d\lambda d\mu}{(2\pi)^2} e^{i\lambda x_1+i\mu(x_2-x_1)}
\langle pS_T|\bar{\psi}_j(0) g(\partial^{\alpha}A^+(\mu n)-\partial^+A^{\alpha}(\mu n))
\psi_i(\lambda n)|pS_T\rangle
\nonumber\\ 
&& \qquad \times
{\partial\over \partial k_2^{\beta}}H^{p}_{ji}
(k_1,k_2)\Bigr|_{k_i=x_i p}\,, \label{der}
\eeq
 where $n^\mu = \delta^\mu_-/p^+$. We recognize the linear part of the field strength tensor $F^{\alpha +}$. The nonlinear part and the Wilson lines (which make the nonlocal operator gauge invariant) will come from the other diagrams in Fig.~1. Taking this for granted, we employ the following parameterization of 
 the resulting nucleon matrix element \cite{Eguchi:2006qz}
\beq
&&\frac{1}{p^+}\int \frac{d\lambda d\mu }{(2\pi)^2} e^{i\lambda x_1+i\mu(x_2-x_1)}\langle pS_T|\bar{\psi}_j(0) gF^{\alpha +}(\mu n)\psi_i(\lambda n)|pS_T\rangle \nonumber \\
&&= \frac{M}{4}(\Slash p)_{ij}\epsilon^{\alpha p n S_T}G_F(x_1,x_2) + i\frac{M}{4}(\gamma_5\Slash p)_{ij} S^\alpha_T \tilde{G}_F(x_1,x_2)\,,
\eeq
 where $M$ is the nucleon mass.\footnote{The relation to the function $T_F(x_1,x_2)$ often used in the literature (e.g., Ref.~\cite{Kouvaris:2006zy}) is 
\beq
G_F(x_1,x_2)=  +\frac{g}{\pi M} T_F(x_1,x_2)\,, \quad T_F(x_1,x_2)= - \int \frac{d\lambda d\mu}{4\pi (p^+)^2}  e^{i\lambda x_1+i\mu(x_2-x_1)}\langle pS_T|\bar{\psi}(0)\gamma^+ \epsilon^{\alpha p n S_T}F_{\alpha}^{\  +}(\mu n)\psi (\lambda n)|pS_T\rangle. 
\eeq
When comparing different definitions in the literature, one has to be careful about the sign convention of the coupling $g$.  } Our conventions are $D^\mu=\partial^\mu-igA^\mu_a t^a$, $\gamma_5=i\gamma^0\gamma^1\gamma^2\gamma^3$ and $\epsilon_{0123}=+1$ so that $\epsilon^{\alpha p n S_T}\equiv \epsilon^{\alpha \lambda \mu\nu}p_\lambda n_\mu S_{T \nu}=- \epsilon^{\alpha\beta}S_{T \beta}$ with $\epsilon^{12}=-\epsilon^{21}=1$. 
  The dimensionless functions $G_F$ and $\tilde{G}_F$ obey the symmetry property
\beq
 G_F(x_1,x_2)=G_F(x_2,x_1)\,, \qquad \tilde{G}_F(x_1,x_2)=-\tilde{G}_F(x_2,x_1)\,. \label{sy}
\eeq


\section{Computation of SSA}

  In this section, we explicitly evaluate (\ref{der})  for the two diagrams in Fig.~2 by computing the derivative of the hard part $\partial H/\partial k$.  The saturation effect is not included, it will be considered in the next section.
 
\subsection{Soft gluonic pole}
\label{pole} 

Let us first calculate the pole part of the left diagram in Fig.~2.  The on-shell conditions for this diagram are 
\beq
(x_1p+k)^2=0\,, \qquad x_2p^\mu +k^\mu -\frac{P_h^\mu}{z}=0\,, \qquad P_h^2\approx 0\,,
\eeq
 where $k^\mu=(k^+=0, k^-=x_g q^-,\vec{k}_T)$. ($q^\mu\approx \delta^\mu_- q^-$ is the target momentum.) The light hadron mass will be neglected. 
It immediately follows that $x_1-x_2=0$, namely, the collinear gluon momentum vanishes. In the literature, this is called the soft gluonic pole (SGP).

At the SGP, the formula (\ref{SFP}) cannot be used. Instead, `master formulas' specific to the SGP have been derived \cite{Kouvaris:2006zy,Koike:2006qv,Koike:2011ns}. However, the diagram under consideration is simple enough and can be computed directly.  
We first note that the color factor for this diagram is $t^bt^at^b=-\frac{1}{2N_c}t^a$, and the function $\tilde{G}_F$ vanishes at the SGP. By convoluting (\ref{der}) with the unpolarized proton/nucleus matrix element (not shown in Fig.~1), we find  
\beq
&& - i\frac{g^2M}{4}  \epsilon^{\alpha\beta}S_{T \beta}\int dx_1dx_2G_F(x_1,x_2)\frac{-1}{2N_c} \int  d^3k \frac{\langle q| A^a_\mu(k)A^a_\nu(-k) |q \rangle}{N_c^2-1} \frac{\partial}{\partial k_2^\alpha} \Biggl\{
\tr [\Slash p \gamma^\nu (\Slash k_2+\Slash k) \Slash p(x_1\Slash p + \Slash k )\gamma^\mu]
\nonumber \\  && \times (-i\pi)(2\pi)^4 \left( \delta((k+x_1p)^2)\delta^{(4)}\left(k_2+k-\frac{P_h}{z}\right) -\delta((k+k_2)^2)\delta^{(4)}\left(x_1p+k-\frac{P_h}{z}\right)  \right) \Biggr\}_{k_2=x_2p} \nonumber \\ 
&& =\frac{2\pi^5 g^2 M}{N_c(N_c^2-1)} \epsilon^{\alpha\beta}S_{T \beta} \int dx_1dx_2 G_F(x_1,x_2)\int  d^3k \langle A_\mu(k)A_\nu(-k)\rangle \label{ba}   \tr [\Slash p \gamma^\nu   \Slash k\Slash p \Slash k \gamma^\mu]  \nn
&&  \qquad \times  \frac{\partial}{\partial k_2^\alpha} \left\{ \delta((k+x_1p)^2)\delta^{(4)}\left(k_2+k-\frac{P_h}{z}\right) - \delta((k+k_2)^2)\delta^{(4)}\left(x_1p+k-\frac{P_h}{z}\right)  \right\}_{k_2=x_2p} \,. \label{del}
\eeq 
[Note that the $k_2^\alpha$-derivative acting on $\Slash k_2$ inside the trace does not contribute due to the property (\ref{sy}).]  We then notice that the gluon field correlator reduces to the unintegrated gluon distribution 
\beq
 \tr [\Slash p \gamma^\mu   \Slash k\Slash p \Slash k \gamma^\nu] \langle A_\mu(k)A_\nu(-k)\rangle &=& 8(p^+)^2\langle -k^2 A^-A^- +k^-A^- k\cdot A + k\cdot A k^-A^--(k^-)^2A_\mu A^\mu \rangle \nonumber \\ 
&=&-8(p^+)^2 \langle F^{-\mu}(k)F^{-}_{\ \ \mu}(-k)\rangle 
\nn
&=& 8(p^+)^2q^- x_gG(x_g,k_T)\,, 
\eeq
  evaluated at $x_g=\frac{k^-}{q^-}= \frac{P_h^-}{zq^-}$. The derivative of the delta functions in the last line of (\ref{del})  should be handled carefully. It is safe to first perform the integrals over $x_1, x_2, k^-, \vec{k}_T$, and then differentiate. We thus obtain  
\beq
\frac{16\pi^5 g^2 M p^+q^-x}{N_c(N_c^2-1)} \epsilon^{\alpha\beta}S_{T \beta} 
\left[ -\frac{1}{k_T^2} \frac{ \partial}{\partial k^\alpha_T}  x_g G(x_g, k_T) G_F(x,x) + \frac{2k_{T\alpha}  }{k_{T}^4} x_g G(x_g,k_{T}) x\frac{d}{dx}G_F(x,x) \right]_{k_T=\frac{P_{hT}}{z}}\,. \label{sg}
\eeq
In this equation, $x = \frac{P_h^+}{zp^+} \approx \frac{x_F}{z}$ where $x_F\equiv \frac{2P^z_{h}}{\sqrt{s}}$ ($s\approx 2p^+q^-$) is the commonly used variable.

\subsection{Soft fermionic pole}
\label{so}

Next we turn to the right diagram in Fig.~2. The on-shell conditions are
\beq
(k+(x_2-x_1)p)^2=0, \qquad  x_2p^\mu+k^\mu-\frac{P_h^\mu}{z}=0\,. \label{right}
\eeq
It follows that $x_1=0$, namely, the incoming quark momentum vanishes. It is thus  called the soft fermionic pole (SFP). The color factor for this diagram is $-if^{abc}t^bt^c = \frac{N_c}{2}t^a$. 
Since $x_1\neq x_2$, we can use (\ref{SFP}) and evaluate (\ref{ba}) as 
\beq
&&i\frac{N_c}{2}\frac{g^2M}{4}\int dx_1dx_2 {\rm P}\frac{1}{x_1-x_2} \int d^3k \frac{\langle A_\mu(k)A_\nu(-k)\rangle}{N_c^2-1} (-i\pi)(2\pi)^4 \nonumber \\
&&\times\Biggl\{ M^{\mu\nu}_{\ \ \alpha}\epsilon^{\alpha pnS_T}G_F(x_1,x_2)\left(\delta((k+(x_2-x_1)p)^2)\delta^{(4)}\left(x_2p+k-\frac{P_h}{z}\right) -(x_1\leftrightarrow x_2)\right) \nonumber \\
&& \qquad +\tilde{M}^{\mu\nu}_{\ \ \alpha}S^\alpha_T \tilde{G}_F(x_1,x_2) \left(\delta((k+(x_2-x_1)p)^2)\delta^{(4)}\left(x_2p+k-\frac{P_h}{z}\right) +(x_1\leftrightarrow x_2)\right)\Biggr\}\,, \label{fin}
\eeq
where
\beq
M^{\mu\nu\alpha}&\equiv &\tr [\Slash p \gamma^\nu (x_2\Slash p+\Slash k)\gamma_\beta] \Bigl(g^{\mu\alpha}(x_2p-k)^\beta - g^{\alpha\beta} (2x_2p+k)^\mu + g^{\beta \mu}(2k+x_2 p)^\alpha\Bigr) \nn 
&= & -4(k^\mu +2x_2p^\mu )(p^\nu k^\alpha -p\cdot k g^{\alpha\nu}) +8 k^\alpha( p^\mu k^\nu + p^\nu k^\mu +2x_2p^\mu p^\nu-p\cdot k g^{\mu\nu})\,,
\label{4}
\eeq
\beq
\tilde{M}^{\mu\nu\alpha}&=& i \tr [\gamma_5\Slash p \gamma^\nu (x_2\Slash p+\Slash k)\gamma_\beta] \Bigl(g^{\mu\alpha}(x_2p-k)^\beta - g^{\alpha\beta} (2x_2p+k)^\mu + g^{\beta \mu}(2k+x_2p)^\alpha\Bigr) \nonumber \\ 
& = &-4p^+  k_\lambda \left( (2x_2p^\mu+k^\mu)\epsilon^{-\alpha\lambda \nu}-2k^\alpha \epsilon^{-\mu\lambda\nu}\right)\,. \label{tilde}
\eeq
In the above, we already used the condition $(k+x_2p)^2=0$ which follows from (\ref{right}) and omitted the terms proportional to $p^\alpha$ since the index $\alpha$ is transverse.

Consider the $G_F$ part. 
The first term in (\ref{4}) becomes, after contracting with $A^\mu A^\nu$, 
\beq
-4 (k^\mu +2x_2p^\mu )(p^\nu k^\alpha -p\cdot k g^{\alpha\nu})A_\mu(k)A_{\nu}(-k)
&=&-\frac{4p^+}{k^-}(k^-k^\mu A_\mu-k^2A^-)(k^\alpha A^--k^- A^\alpha) \nonumber \\
&=&\frac{4p^+k_\mu}{k^-}F^{-\mu}F^{-\alpha}\,.
\eeq
We then use, for $\alpha,\beta$ transverse,  
\beq
\frac{1}{q^-}\langle F^{-\alpha}F^{-\beta}\rangle = \frac{1}{2}\delta^{\alpha\beta}x_gG(x_g,k_T ) + \frac{1}{2}\left(\frac{2k^\alpha k^\beta}{k_T^2}-\delta^{\alpha\beta}\right)x_gh(x_g,k_T) = \frac{k^\alpha k^\beta}{k_T^2} x_gG(x_g,k_T) \,,
\eeq
where $h(x_g,k_T)$ is the so-called linearly polarized gluon distribution, and in the last equality we used the fact that $G(x_g,k_T)=h(x_g,k_T)$  at small-$x$ in the present approximation \cite{Metz:2011wb,Dominguez:2011br}. 
Adding the second term in (\ref{4}), we find
\beq
M^{\mu\nu\alpha}\langle A_\mu(k)A_\nu(-k)\rangle = \frac{4p^+q^- k^\alpha}{k^-}x_gG(x_g,k_T)\,.
\eeq
In the $\tilde{G}_F$ part, we drop the second term in (\ref{tilde}) which is antisymmetric in $\mu$ and $\nu$. The first term gives
\beq
-4p^+ k_\lambda (2x_2p^\mu+k^\mu)\epsilon^{-\alpha\lambda \nu} \langle A_\mu(k)A_\nu(-k)\rangle =-4p^+\epsilon^{\beta\alpha}  \langle (2x_2p^+A^-+k^\mu A_\mu)( k_\beta A^- - k^- A_\beta)\rangle \nonumber \\
=4p^+\epsilon^{\beta\alpha}  \frac{k^\mu}{k^-}\langle F^-_{\ \ \mu}F^-_{\ \ \beta}\rangle =-\frac{4p^+q^-}{k^-}\epsilon^{\beta\alpha}k_\beta x_gG(x_g,k_T)\,.
\eeq


(\ref{fin}) therefore becomes 
\beq 
&&-\frac{16\pi^5g^2 N_c M}{N_c^2-1}  \frac{q^-p^+ z^3  }{P_{hT}^4} x_g G(x_g,P_h/z) \epsilon^{\alpha\beta}P_{h\alpha}S_{T \beta} \int \frac{dx_1dx_2}{x_1-x_2}  \nonumber \\ 
&&\quad \times
\left\{G_F(x_1,x_2)\Bigl(x_2^2\delta(x_1)\delta(x_2-x)-(x_1\leftrightarrow x_2)\Bigr) +\tilde{G}_F(x_1,x_2)\Bigl(x_2^2\delta(x_1)\delta(x_2-x)+(x_1\leftrightarrow x_2) \Bigr)\right\} \nonumber \\ 
&&= \frac{32\pi^5 g^2 N_c M }{N_c^2-1}\frac{q^-p^+ z^3 x}{P_{hT}^4} x_g G(x_g,P_h/z) \epsilon^{\alpha\beta}P_{h \alpha}S_{T \beta}  \bigl(G_F(0,x)+\tilde{G}_F(0,x)\bigr)\,. \label{sf}
\eeq

\subsection{Spin-dependent cross section and matching to the collinear result}

To obtain the spin-dependent cross section, we add (\ref{sg}) and (\ref{sf}) and multiply by 
\beq
\frac{1}{2s}\frac{dP_h^+d^2P_{hT}}{(2\pi)^3 2P_h^+} \int \frac{dz}{z^2}D(z)\,,
\eeq
 where $D$ is the fragmentation function. 
The result is 
\beq
\frac{d\sigma}{dy_h d^2P_{hT}} &=& \frac{\pi^2 g^2 Mx_F}{4N_c(N_c^2-1)} \epsilon^{\alpha\beta}S_{T\beta} \int_{x_F}^1 \frac{dz}{z^3}D(z) \Biggl\{ -  \frac{1}{(P_{hT}/z)^2} \frac{\partial}{\partial P_h^\alpha/z} x_g G(x_g,P_{hT}/z) G_F(x,x)  
\nonumber \\ 
&& \qquad+ \frac{ 2P_{h\alpha}/z}{(P_{hT}/z)^4}x_g G(x_g,P_{hT}/z) \left(x \frac{d}{dx}G_F(x,x) +N_c^2(G_F(0,x)+\tilde{G}_F(0,x))\right)\Biggr\} \,, \label{final}
\eeq
appropriate for the kinematic region $x_F\sim {\mathcal O}(1)$ where 
\beq
x= \frac{x_F}{z} \sim {\mathcal O}(1)\,, \qquad  x_g=\frac{P_{hT}^2}{szx_F} \ll 1\,.
\eeq

Let us check that (\ref{final}) matches the known result obtained within the collinear factorization approach relevant at high-$P_T$.  
At large-$k_T$,  $x_g G(x_g,k_T)\sim 1/k_T^2$, and in this regime (\ref{final}) takes the form
\beq
\frac{d\sigma}{dy_h d^2P_{hT}} &\approx & -\frac{\pi^2 g^2 M}{2N_c(N_c^2-1)} \epsilon^{\alpha\beta}P_{h\alpha} S_{T\beta} \int \frac{dz}{z^3}D(z) 
\nonumber \\ 
&&  \times\Biggl\{  \frac{x_g G(x_g,P_{hT}/z) }{(P_{hT}/z)^4} x\left(G_F(x,x) -x \frac{d}{dx}G_F(x,x) -N_c^2(G_F(0,x)+\tilde{G}_F(0,x))\right)\Biggr\} \,.  \label{col2}
\eeq
On the other hand, the contribution from the SGP in the collinear approach is \cite{Kouvaris:2006zy,Koike:2007rq}  
\beq
\frac{d\sigma^{SGP}}{dy_h d^2P_{hT}}  = - \frac{\pi M \alpha_s^2}{s} \epsilon^{\alpha\beta}P_{h\alpha} S_{T\beta}  \int \frac{dz}{z^3} D(z)\int \frac{dx'}{x'}G(x')   \frac{1}{\hat{u}^2} \left(G_F(x,x)-x\frac{d}{dx}G_F(x,x)\right)
 \sigma_{qg\to q}\,,
\label{col}
\eeq
 where $G(x')$ is the collinear (integrated) gluon distribution and $\hat{s},\hat{t},\hat{u}$ are the Mandelstam variables at the partonic level ($\hat{s}=xx' s$, etc.). In (\ref{col}) we have kept only one partonic subprocess $qg\to qg$ with the gluon in the final state being unobserved. In the forward region, this should be the dominant channel. The corresponding cross section receives contributions from both the initial ($I$) and final ($F$) state interactions 
\beq
\sigma_{qg\to q} = \sigma^I + \sigma^F\left(1+\frac{\hat{u}}{\hat{t}} \right)\,,
\eeq
 where 
\beq
\sigma^I = \frac{1}{2(N_c^2-1)} \left(\frac{\hat{s}}{\hat{u}} + \frac{\hat{u}}{\hat{s}}\right) \left(1-N_c^2\frac{\hat{u}^2}{\hat{t}^2} \right)\,,  \qquad 
\sigma^F = \frac{1}{2N_c^2(N_c^2-1)}  \left(\frac{\hat{s}}{\hat{u}} + \frac{\hat{u}}{\hat{s}}\right)
\left(1+2N_c^2\frac{\hat{s
}\hat{u}}{\hat{t}^2}\right)\,.
\eeq
We see that, in the forward region where $\hat{s}\approx -\hat{u} \gg |\hat{t}|$, $\sigma^F$ is enhanced by a kinematic factor $\hat{u}/\hat{t} \gg 1$. We thus neglect $\sigma^I$ and approximate as 
\beq
\sigma_{qg\to q} \approx \frac{\hat{u}}{\hat{t}} \sigma^F \approx \frac{1}{N_c^2-1} 
\frac{2\hat{s}^3}{-\hat{t}^3} = \frac{1}{N_c^2-1} \frac{2\hat{s}^3}{k_T^6}\,. \label{sigma}
\eeq
Returning to (\ref{col2}), at large-$k_T$ we can use the relation 
\beq
G(x_g,k_T) \approx \frac{\alpha_s}{2\pi^2} \frac{1}{k_T^2} \int \frac{dx'}{x'} G(x') 
P_{gg}(x_g/x') + \cdots\,.
\eeq
 where $P_{gg}$ is the gluon splitting function. 
At small-$x$ we may approximate  $P_{gg}(z) \approx 2N_c/z$. Substituting this into (\ref{col2}) and comparing the result with (\ref{col}), we find that they agree. Similarly, 
the contribution from the SFP in the collinear framework is given by \cite{Koike:2009ge}
\beq
\frac{d\sigma^{SFP}}{dy_h d^2P_{hT}}  = - \frac{\pi M \alpha_s^2}{2s} \epsilon^{\alpha\beta}P_{h\alpha} S_{T\beta}  \int \frac{dz}{z^3} D(z)\int \frac{dx'}{x'}G(x')   \frac{1}{-\hat{u}} \left(G_F(0,x)+ \tilde{G}_F(0,x)\right)
 \tilde{\sigma}_{qg\to q}\,, \label{koike}
\eeq
 where again we only picked up the channel $qg\to qg$. To the order of interest, 
\beq
\tilde{\sigma}_{qg\to q} \approx - \frac{4N_c^2}{N_c^2-1}\frac{\hat{s}^2}{k_T^6}\,.
\eeq
It is easy to check that in this approximation (\ref{koike})  agrees with the SFP part of (\ref{col2}). Actually, in the collinear calculation \cite{Koike:2009ge}  there is not a clean separation between initial and final state interactions for the SFP contribution. We are however inclined to interpret our result as coming from the initial state interaction, see the right diagram in Fig.~2.

We have thus seen that (\ref{final}) correctly reproduces the dominant part of the fully collinear results  in the forward region at high-$P_{hT}$. The formula can be used for smaller values of $P_{hT}$ (around a few GeV), but eventually we have the constraint $P_{hT}\gg \Lambda_{QCD}$ because we have performed the collinear expansion on the projectile side.

\section{Including the saturation effect}
\label{saturation}

By construction, the formula (\ref{final}) has been obtained in the two-gluon exchange (leading twist) approximation. At small-$x_g$ such that $\alpha_s \ln 1/x_g\sim {\mathcal O}(1)$,  one can consistently include the BFKL evolution effects in the unintegrated gluon distribution  $G(x_g,k_{T})$. However, the two-gluon approximation breaks down when the gluon saturation (multiple scattering or higher twist) effect becomes important. This inevitably happens for very small values of $x_g$ and/or for a heavy nucleus target. In the saturated regime, a new parturbative scale, the so-called saturation momentum $Q_s(x_g)$ is dynamically generated \cite{Iancu:2003xm}, and  the particle production around $P_{hT}\sim Q_s$ is significantly modified from the leading-twist result.  We now  discuss how to generalize (\ref{final}) in the saturation environment. 
 
The multiple scattering of the collinear quark can be resummed to all orders via the eikonal approximation. This  effectively converts the quark-gluon vertex into a Wilson line in the fundamental representation 
\beq
ig\gamma^\mu A^a_\mu (k)t^a \to \gamma^+ \int \frac{d^2\vec{x}}{(2\pi)^3} e^{i\vec{x}\cdot \vec{k}_T} (U(\vec{x})-1) \,, \qquad U(\vec{x})=\exp\left(ig\int dx^+ A_a^-(x^+,\vec{x})t^a \right)\,,
\eeq
where $k^- (\ll k_T)$ is neglected. Similarly, the interaction of the collinear gluon with the target (see the right diagram of Fig.~2) can be promoted to  a Wilson line in the adjoint representation $\tilde{U}_{ab}(\vec{z})-\delta_{ab}$. There is, however, a caveat here. If one naively applies the eikonal approximation to the three-gluon vertex in (\ref{4}), one only keeps the term $\sim g^{\alpha\beta} p^\mu$ with the index $\alpha$ being transverse \cite{Forshaw:1997dc}. This describes a transversely polarized gluon and leads to the term $-8x_2p^\mu p^\nu k^\alpha$ in the second line of (\ref{4}).   However, one cannot neglect the term $\sim g^{\beta\mu}k^\alpha$ which involves a longitudinally polarized gluon, because it actually gives  a larger contribution $+16x_2p^\mu p^\nu k^\alpha$ with an opposite sign. This problem was  previously encountered in the context of SSA in direct photon production and Drell-Yan  \cite{Schafer:2014zea,Zhou:2015ima}. There the authors employed an elaborate  formalism of gluon production in the covariant gauge developed  in \cite{Blaizot:2004wu}.  Our task here is simpler, since there is a strong constraint that the formula (\ref{final}) must be recovered in the `dilute' limit. Knowing this, we can arrive at the desired result via the following sequence of observations. 

Let us first consider the color structure. Once we include the multiple scattering, the two diagrams in Fig.~2 can be treated at the same time. Working in the coordinate space, to the left side of the cut we assign the Wilson lines as 
\beq
 U(\vec{x})(\tilde{U}(\vec{z})-1)_{ba} + (U(\vec{x})-1) \tilde{U}_{ba}(\vec{z}) -(U(\vec{x})-1)(\tilde{U}(\vec{z})-1)_{ba} = U(\vec{x})\tilde{U}_{ba}(\vec{z}) -\delta_{ab}\,.
\eeq
The last term on the left hand side subtracts the double counting.  Then the overall  color structure for the diagrams in Fig.~2 is 
\beq
&&\left\langle (U^\dagger(\vec{y})-1) t^b  (U(\vec{x})\tilde{U}_{ba}(\vec{z}) -\delta_{ab}) \right\rangle  = \frac{2}{N_c^2-1} \left\langle {\rm Tr} \left[(U^\dagger(\vec{y})-1)   (U(\vec{z})t^bU^\dagger(\vec{z}) U(\vec{x})-t^b)t^b\right]  \right\rangle t^a  \nn
&& =\frac{2}{N_c^2-1} \biggl\langle\frac{1}{2} {\rm Tr} [U^\dagger(\vec{y})U(\vec{z})] {\rm Tr}[U^\dagger(\vec{z})U(\vec{x})]  -\frac{1}{2N_c} {\rm Tr} [ U^\dagger(\vec{y})U(\vec{x})]  -C_F  {\rm Tr} [U^\dagger(\vec{y})]  +\frac{1}{2N_c}  {\rm Tr}[ U(\vec{x})] \nn
&&  \qquad \qquad \qquad -\frac{1}{2}{\rm Tr}[U(\vec{z})] {\rm Tr}[U^\dagger(\vec{z})U(\vec{x})]+  C_F N_c \biggr\rangle t^a\,, \label{two}
\eeq
 where  $C_F=\frac{N_c^2-1}{2N_c}$ and $\vec{y}$ is the quark coordinate in the complex-conjugate amplitude. We shall only keep the first two terms in (\ref{two}) :  $\langle {\rm Tr}[U(\vec{x})]\rangle$ represents the S-matrix of a single quark. This vanishes due to infrared divergences.\footnote{The operator ${\rm Tr} U(\vec{x})$ is not gauge invariant (in the sense of \cite{Hatta:2005as}), and the small-$x$ evolution equation of such non-gange-invariant operators contain infrared divergences. Thus the expectation value $\langle {\rm Tr}[U(\vec{x})]\rangle$, even if it is nonzero in simple models, immediately goes to zero once the quantum evolution effects are included.  } The term $\langle {\rm Tr}[U(\vec{z})] {\rm Tr}[U^\dagger(\vec{z})U(\vec{x})]\rangle$ is independent of  $\vec{y}$, so it corresponds to the case $P_{hT}=0$ which can be discarded (see below). 


We now go to the momentum space by introducing notations   $\ell_T$ and $k_T$ for the momenta transferred to the collinear quark and gluon, respectively.\footnote{We thus use a different notation $k \to \ell$ from Section \ref{pole}. } Using the large-$N_c$ approximation
\beq
\langle q| {\rm Tr} [U^\dagger(\vec{y})U(\vec{z})] {\rm Tr}[U^\dagger(\vec{z})U(\vec{x})] |q\rangle \approx \frac{\langle q| {\rm Tr} [U^\dagger(\vec{y})U(\vec{z})] |q\rangle\langle q| {\rm Tr}[U^\dagger(\vec{z})U(\vec{x})] |q\rangle}{\langle q|q\rangle},  \label{nc}
\eeq
where $\langle q|q\rangle = 2q^- (2\pi)^3\delta^{(3)}(0) = 2q^- \int dx^+ d^2\vec{x}$, 
we find 
\beq
&&  \int \frac{d^2\vec{x} d^2\vec{y} d^2\vec{z}}{(2\pi)^6} \, (2\pi)^2\delta(\vec{k}_T+\vec{\ell}_T-\vec{P}_{hT}/z) e^{i\vec{k}_T\cdot \vec{z}+i\vec{\ell}_T \cdot \vec{x} -i\frac{\vec{P}_{hT}}{z}\cdot \vec{y}}
\nn 
&& \qquad \times \left\langle {\rm Tr} [U^\dagger(\vec{y})U(\vec{z})] {\rm Tr}[U^\dagger(\vec{z})U(\vec{x})]  -\frac{1}{N_c} {\rm Tr} [ U^\dagger(\vec{y})U(\vec{x})] \right\rangle
 \nn 
&&\approx  \langle q|q\rangle \delta(\vec{k}_T+\vec{\ell}_T-\vec{P}_{hT}/z) \left(  \frac{N_c^2}{ \int d^2\vec{x}}  F(x_g,\ell_T) F(x_g, P_{hT}/z) -\delta^{(2)}(\vec{k}_T) F(x_g,P_{hT}/z)\right) \,. \label{nis}
\eeq
 Here, $F$ is defined as the Fourier transform of the dipole S-matrix 
\beq
F(x_g,k_T) \equiv \int \frac{d^2\vec{x} d^2\vec{y}}{(2\pi)^2} e^{i\vec{k}_T\cdot (\vec{x}-\vec{y})}  \frac{\langle q|  \frac{1}{N_c}{\rm Tr} [U^\dagger(\vec{y})U (\vec{x})] |q\rangle }{\langle q|q\rangle}\,. \label{four}
\eeq 

The second term in (\ref{nis}) is the direct generalization of the left diagram of Fig.~2. Since there is no momentum transfer to the collinear gluon $(k_T=0)$, the kinematics that determines the position of the pole is unchanged, namely, the SGP at $x_1=x_2$ survives.  
 We can then immediately write down a contribution to SSA 
\beq
\frac{d\sigma^{SGP}}{dy_h d^2P_{hT}} &=& \frac{\pi Mx_F}{2(N_c^2-1)} \epsilon^{\alpha\beta}S_{T\beta} \int_{x_F}^1 \frac{dz}{z^3}D(z) \Biggl\{ - \frac{1}{(P_{hT}/z)^2} \frac{\partial}{\partial P_h^\alpha/z} \left( \frac{P^2_{hT}}{z^2} F(x_g,P_{hT}/z) \right) G_F(x,x)  
\nonumber \\ 
&& \qquad+ \frac{ 2P_{h\alpha}/z}{(P_{hT}/z)^2} F(x_g,P_{hT}/z) x \frac{d}{dx}G_F(x,x) \Biggr\} \,. \label{sat}
\eeq
 It is known that $k_T^2 F(x_g, k_T)$ is a suitable generalization of the unintegrated gluon distribution in the presence of  saturation. We thus see that, in the SGP sector, the net effect of multiple scattering is simply to replace
\beq
\frac{x_g G(x_g,k_T)}{k_T^2} \to \frac{N_c}{2\pi^2 \alpha_s} F(x_g,k_T)\,, \label{norm}
 \eeq 
in the corresponding part of the formula (\ref{final}). 
The normalization factor in (\ref{norm}) agrees with the one given in \cite{Dominguez:2011wm}. 

We now turn to the first term of (\ref{nis}) which is nonlinear in the gluon density. Naively, we expect that this term represents  the generalization of the SFP in the saturation environment. Surprisingly, however, it turns out that the coefficient of this term identically vanishes in the presence of nonvanishing momentum transfer $\ell$.    
To show this, we first note that there are now two on-shell conditions 
\beq
( (x_2-x_1)p+k)^2=0 \,, \quad (x_1p + \ell)^2=0\,. \label{only}
\eeq
   The first condition is the same as in (\ref{right}) and the second condition effectively comes from the $\ell^-$-integration. 
Together with the momentum conserving delta function $\delta^{(4)} (x_2p+k+\ell -P_h/z)$, the only solution to (\ref{only}) is, with $ \ell^+=k^+=0$, 
\beq 
 x_1=\beta x_2\,, \quad \ell^\mu = \beta \frac{P_h^\mu}{z} \,, \quad k^\mu=(1-\beta) \frac{P^\mu_h}{z} \,, \quad  (\mu \neq +)  \label{com}
\eeq
 for $0\le\beta \le 1$. One can then write 
\beq
\delta\left(((x_2-x_1)p+k)^2\right) = \delta\left(  (1-\beta) \frac{P^2_{hT}}{z^2} -\frac{1-\beta}{\beta}\ell_T^2 - k_T^2\right) = \delta \left(\beta \left(\frac{\vec{P}_{hT}}{z} -\frac{\vec{\ell}_T}{\beta}\right)^2\right)\,. 
\eeq 
 $\beta=0$ corresponds to a SFP, while $\beta=1$ corresponds to a SGP. In between, there is a continuum of poles for different values of $\beta$ and one has to integrate over $\beta$. 

 Next we look at the `hard part'. As mentioned above, the correct approximation to the three-gluon vertex in (\ref{4}) is\footnote{In \cite{Schafer:2014zea}, this structure is hidden in the effective vertex $C_U^\mu$ introduced in \cite{Blaizot:2004wu}
\beq
\Slash C_U((x_2-x_1)p+k,p_T) \approx  \frac{p_{T\alpha}}{ (x_2-x_1)p^+ + i\epsilon}   \left( - 2(x_2-x_1)p^+ \gamma^\alpha + 2 k^\alpha \gamma^+ \right)\,,
\eeq
 where $p_T$ is the transverse momentum of the collinear gluon which can be identified with $k_{2T}$ in Section \ref{co}. 
} 
\beq
\gamma_\beta (-g^{\alpha\beta}2(x_2-x_1) p^\mu + 2k^\alpha g^{\beta \mu} ) \approx \delta^{\mu +} ( - 2(x_2-x_1)p^+ \gamma^\alpha + 2k^\alpha \gamma^+) \,, \label{three}
\eeq
 where we reinstated  $x_1$, since $x_1$ no longer vanishes in general (see (\ref{com})). Then 
the trace calculation in (\ref{4}) becomes 
\beq
&& \tr \left[\Slash p \gamma^+ (x_2\Slash p+\Slash k+\Slash \ell)\gamma_\beta \frac{ x_1\Slash p + \Slash \ell}{2x_1p^+} \gamma^+\right] \Bigl(- 2g^{\alpha\beta}(x_2-x_1)p^+ + 2g^{\beta +}k^\alpha\Bigr) \nn 
&&=(p^+)^2 \left( -8(x_2-x_1)  (k+\ell)^\alpha + 16k^\alpha x_2  -8 \frac{x_2}{x_1}(x_2-x_1) \ell^\alpha \right) \nn
&&= 8(p^+)^2(x_1+x_2) \left(\frac{P_{hT}^\alpha}{z} - \frac{x_2}{x_1}\ell^\alpha\right) \,. \label{be}
\eeq
After inserting the solution (\ref{com}), we find that (\ref{be}) vanishes identically, for any value of $\beta$. Similarly, it is easy to check that the generalization of (\ref{tilde}) to the case $\ell \neq 0$ also vanishes when evaluated at the solution (\ref{com}).  Then how can one recover the result in the previous section? The answer is that (\ref{be}) gives a finite contribution if it is multiplied by a singular function. This is indeed the case for the second term in (\ref{nis}) which contains a delta function singularity $\delta^{(2)}(\vec{k}_T)$, and therefore leads to a finite result (\ref{sat}). However this does not happen for the first term in (\ref{nis}), as long as the function $F(\ell_T=\beta P_{hT}/z)$ has a smooth behavior as one would  expect in the saturation regime.\footnote{This observation does not rely on the large-$N_c$ approximation (\ref{nc}). Even at finite-$N_c$, the first term in (\ref{nis}) defines a smooth function of $\ell_T$. }  Only when one assumes the form
\beq
F(\ell_T) = \delta^{(2)}(\vec{\ell}_T) \int d^2\vec{x}\,,
\eeq
 does one get a finite contribution and thereby recover the SFP contribution in (\ref{final}).  

We have also cross-checked the above results by working in the light-cone gauge $A^+=0$ for the polarized proton.  
By using the principal-value 
prescription for the spurious pole $1/k^+$ in the light-cone gauge propagator, we can avoid a potential phase  from this pole. We then evaluated the same
set of diagrams as in the covariant gauge calculations above, where the initial and final state interaction
effects generate the necessary phase for a non-zero SSA. 

We thus conclude that the SFP disappears in the saturation environment, and therefore, (\ref{sat}) is our final result. The formula is valid in the forward region $x_F={\mathcal O}(1)$ and for $P_{hT} \gg \Lambda_{QCD}$. In particular, the formula is most relevant and phenomenologically useful around the saturation momentum $P_{hT}\sim Q_s(x_g)\gg \Lambda_{QCD}$ where the function $P_{hT}^2 F(P_{hT})$ has a maximum. 


\section{Discussions} 
 
In this paper, we have computed the spin-dependent cross section in light-hadron production $p^\uparrow A \to h X$ including the saturation effect in the target. The use of the hybrid approach allows us to not only check the consistency with the fully collinear calculations in the literature, but also explicitly study the fate of the soft gluonic and soft fermionic poles in the saturation environment.  We have shown that leading terms in the forward region come from the SGP associated with the final state interaction, whereas the SFP is washed out by the saturation effect. From our viewpoint, the way the SFP is recovered in the dilute limit is rather nontrivial. 

 We have limited our discussions to the collinear twist-three functions for the polarized proton. Much of our 
derivations can be extended to the Collins contributions, i.e., taking
into account the collinear twist-three fragmentation functions for the final 
state hadrons, instead of the $k_T$-dependent fragmentation function as in  \cite{Kang:2011ni}. Again, a `hybrid approach' can be formulated,
and similar results shall be obtained. We leave that for a future 
publication. 

We  also note that there has been a debate over the sign mismatch 
in the twist-three function $G_F$ extracted from  SSAs in 
inclusive hadron production and 
semi-inclusive deep inelastic scattering (SIDIS)~\cite{Kang:2011hk}. 
To nail down this issue, a comprehensive analysis 
of all available experimental data is greatly needed. Our formula (\ref{sat}) can offer a  relatively clean environment to access the information about the sign of $G_F(x,x)$.   

Finally, we conclude this paper with  phenomenological implications of our result on the experimentally measured asymmetry
\beq
A_N=\frac{d\sigma^\uparrow -d\sigma^\downarrow}{d\sigma^\uparrow + d\sigma^\downarrow}\,. \label{an}
\eeq 
Here we focus on the dependence of $A_N$ on the mass number $A$ of the target nucleus,   which has been recently studied by the STAR collaboration at RHIC \cite{Heppelmann:2016siw}. 

In our approach, the $A$-dependence comes from that of the saturation momentum $Q_{sA}^2 \sim A^{1/3}$. 
  In the forward region $x_F \approx 1$, it is expected that the dominant term in (\ref{sat})  is the derivative term $x\frac{d}{dx}G_F$ \cite{Qiu:1998ia}. 
 Since this term is proportional to $F$  (and not the derivative of $F$), the nuclear effects contained in $F$ cancel in the ratio (\ref{an}). We thus find that SSA is independent of  $A$ 
\beq
 \frac{A_N^{pA}}{A_N^{pp}} \sim 1\,, \label{22}
\eeq
This holds as long as the formula (\ref{sat}) is valid, namely, for $P_{hT}\gg \Lambda_{QCD}$. (\ref{22}) appears to be consistent with the preliminary STAR data \cite{Heppelmann:2016siw}. 

Let us contrast this result with other arguments.  A phenomenological study based on the $k_T$-factorization  \cite{Boer:2006rj} gives a formula that is sensitive to the derivative of $F$, $d  \sigma^\uparrow -d\sigma^\downarrow \sim \partial F/\partial P_{hT}$. It is thus similar to the first term in (\ref{sat}). 
If we assume the form $F(k_T)\sim e^{-k_T^2/Q_s^2}$ at low momentum, the derivative brings down the factor $1/Q_s^2$ so that 
\beq
\frac{A_N^{pA}}{A_N^{pp}} \sim \frac{Q_{sp}^2}{Q_{sA}^2} \sim \frac{1}{A^{1/3}} <1\,. \quad (P_{hT}\lesssim Q_s)
 \label{11}
\eeq   
For the gold nucleus, this means a significant suppression $1/A^{1/3} \approx 0.17$.  
 On the other hand, the contribution from the Collins fragmentation function alone \cite{Kang:2011ni} shows the same behavior as (\ref{22}) for   $P_{hT} \gg  Q_s$ and (\ref{11}) for $P_{hT}\ll Q_s$. 
 The ongoing experiment at RHIC can further test the different behaviors (\ref{22}) and (\ref{11}), and thereby help clarify the origin of SSA in the forward region.  
We hope that such an analysis is available soon.

\acknowledgements
Y.~H. thanks Edmond Iancu for a helpful conversation. He also thanks the Center of Nuclear Matter Science of Central China Normal University for hospitality during the early stage of this work. This material is based upon work supported by the U.S. Department of Energy, 
Office of Science, Office of Nuclear Physics, under contract number 
DE-AC02-05CH11231, and by the NSFC under Grant No.~11575070.

\end{document}